\documentclass[conference]{IEEEtran}
\IEEEoverridecommandlockouts
\usepackage{cite}
\usepackage{amsmath,amssymb,amsfonts}
\usepackage{algorithm}
\usepackage{graphicx}
\usepackage{textcomp}
\usepackage{hyperref}
\hypersetup{hidelinks}
\usepackage{tikz}
\usepackage{svg}
\makeatletter
\let\MYcaption\@makecaption
\makeatother
\usepackage[font=footnotesize, skip=0pt]{caption}
\usepackage[font=footnotesize, skip=0pt]{subcaption}
\makeatletter
\let\@makecaption\MYcaption
\makeatother
\usepackage{xcolor}
\usepackage{color, colortbl}
\usepackage{listings}
\usepackage{multirow}
\usepackage{hhline}
\usepackage{array}
\usepackage{cite}
\usepackage{amsmath,amssymb,amsfonts}
\usepackage{tabularx}
\usepackage{rotating}
\usepackage{makecell}

\usepackage{algpseudocode}
\usepackage{booktabs}
\usepackage{ragged2e} 
\usepackage{pifont}

\newcommand{\cmark}{\ding{51}} 
\newcommand{\xmark}{\ding{55}} 



\newif\ifarxiv

\arxivtrue

\def\BibTeX{{\rm B\kern-.05em{\sc i\kern-.025em b}\kern-.08em
    T\kern-.1667em\lower.7ex\hbox{E}\kern-.125emX}}

\ifarxiv

\else

\fi

\definecolor{seg_color}{HTML}{1f77b4}
\definecolor{segrr_color}{HTML}{ff7f0e}
\definecolor{hyb_color}{HTML}{2ca02c}

\definecolor{mygreen}{HTML}{008A00}
\definecolor{myred}{HTML}{A20025}
\definecolor{myblue}{HTML}{0050EF}

\newcommand{\redcirclednum}[1]{%
  \tikz[baseline=(char.base)]{
    \node[shape=circle,
          fill=myred,
          text=white,
          inner sep=0pt,
          minimum size=1.2em,
          font=\bfseries,
          align=center] (char) {#1};
  }%
}

\newcommand{\greencirclednum}[1]{%
  \tikz[baseline=(char.base)]{
    \node[shape=circle,
          fill=mygreen,
          text=white,
          inner sep=0pt,
          minimum size=1.2em,
          font=\bfseries,
          align=center] (char) {#1};
  }%
}

\newcommand{\bluecirclednum}[1]{%
  \tikz[baseline=(char.base)]{
    \node[shape=circle,
          fill=myblue,
          text=white,
          inner sep=0pt,
          minimum size=1.2em,
          font=\bfseries,
          align=center] (char) {#1};
  }%
}

\newif\ifshowinternalcomments
\showinternalcommentstrue
\showinternalcommentsfalse
\newcommand{\internalcomment}[1]{\ifshowinternalcomments\textcolor{red}{\textbf{[Internal clarification:} #1\textbf{]}}\fi}

\newif\ifreview

\reviewfalse

\ifreview
\newcommand{\reviewtxt}[1]{\textcolor{blue}{#1}}
\else
\newcommand{\reviewtxt}[1]{\textcolor{black}{#1}}
\fi

\ifreview
\newcommand{\reviewtxtv}[1]{\textcolor{blue}{#1}}
\else
\newcommand{\reviewtxtv}[1]{\textcolor{black}{#1}}
\fi

\definecolor{plt_1_color}{HTML}{1F77B4}
\definecolor {plt_2_color}{HTML}{FF7F0E}
    
\definecolor{idegreen}{RGB}{34,139,34} 
\newcommand{\IDEComment}[1]{\hfill\small{\ttfamily\color{idegreen}\Comment{#1}}}

\title{MEDEM: \underline{M}ulti-\underline{E}ngine DL Accelerator \underline{DE}sign \underline{M}ethodology}

\begin{document}

\author{
\IEEEauthorblockN{Fareed Qararyah \href{https://orcid.org/0000-0002-3955-2836}{\includegraphics[width=1em]{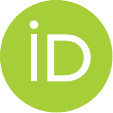}}, Mohammad Ali Maleki \href{https://orcid.org/0000-0002-9019-3605}{\includegraphics[width=1em]{figures/orcid_logo.pdf}}, Pedro Trancoso \href{https://orcid.org/0000-0002-2776-9253}{\includegraphics[width=1em]{figures/orcid_logo.pdf}}}
\IEEEauthorblockA{\textit{Department of Computer Science and Engineering} \\
\textit{Chalmers University of Technology and University of Gothenburg, Gothenburg, Sweden}\\
\{qarayah, mohammad.ali.maleki, ppedro\}@chalmers.se}
\ifarxiv
\thanks{This work was supported by the Swedish Foundation for Strategic Research (contract number CHI19-0048) under the PRIDE project and the European High-Performance Computing Joint Undertaking (JU) under Framework Partnership Agreement No 800928 and Specific Grant Agreement No 101036168 (EPI SGA2).
The JU is supported by the European Union’s Horizon 2020 research and innovation program and by Croatia, France, Germany, Greece, Italy, Netherlands, Portugal, Spain, Sweden, and Switzerland. This work was also partially funded by the Swedish Foundation for Strategic Research under the SSF-FuSS QuantumStack project (grant number FUS21-0063).}
\fi
}

\maketitle

\thispagestyle{plain}
\pagestyle{plain}

\begin{abstract}
Multi-engine deep learning (DL) accelerators are becoming increasingly prevalent as they address the heterogeneity and growing complexity of modern DL workloads. To efficiently process diverse DL workloads, these accelerators must incorporate combinations of engines with complementary capabilities to match the distinct computational characteristics of these workloads' heterogeneous kernels. However, existing multi-engine DL accelerator design approaches lack a systematic methodology, leaving fundamental questions unresolved. These include how to co-design engines for workloads with diverse computational characteristics and which engine combinations minimize aggregate execution costs (such as time or energy) across such workloads. Addressing these questions requires efficient exploration of exponentially large design spaces.

To address these questions systematically, this work proposes \underline{M}ulti-\underline{E}ngine DL accelerator \underline{DE}sign \underline{M}ethodology (MEDEM). MEDEM defines generic engine abstractions, co-designs candidate instances (engines), and selects a combination of co-designed engines to minimize aggregate execution cost given diverse DL workloads and a resource budget. MEDEM encompasses a set of design strategies that efficiently navigate the exponentially large design spaces of engine co-design and combination selection, identifying highly optimized multi-engine accelerators. A comprehensive evaluation demonstrates that MEDEM identifies accelerators that outperform state-of-the-art designs, delivering geometric-mean improvements of up to $4.84\times$ in energy-delay product (EDP) and $1.59\times$ in throughput. The improvements are achieved using different resource budgets, demonstrating MEDEM's scalability, and using 51 single- and multi-model DL workloads, demonstrating its generalizability.

\end{abstract}


\section{Introduction}

As monolithic Deep Learning (DL) accelerators face diminishing returns from further optimization, multi-engine accelerators offer a more scalable alternative and open up new design opportunities~\cite{verhelst2022ml, gao2019tangram, zheng2022atomic, kwon2021heterogeneous, gong2025crane}. Hence, they are increasingly adopted both in research and in industry. Representative research-driven multi-engine accelerators include Simba~\cite{shao2021simba}, Tangram~\cite{gao2019tangram}, Maelstrom~\cite{kwon2021heterogeneous}, and Gemini~\cite{cai2024gemini}. Representative accelerators widely deployed include Meta's MTIA~\cite{coburn2025meta}, Tesla’s Dojo~\cite{talpes2023microarchitecture}, and Google’s TPU v4\footnote{The TPU architecture evolved from a single systolic array in v1 to multiple TensorCores, each integrating four matrix multiplicative units (MXUs), in v4.}~\cite{jouppi2023tpu}.

\begin{figure}[t]
     \centering
     \includegraphics[width=\columnwidth]{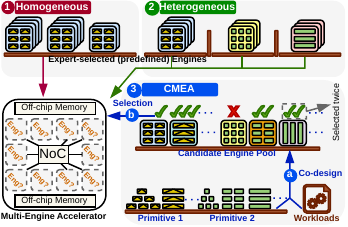}
     \caption{Engine design alternatives in multi-engine DL accelerators: \protect \redcirclednum{1} homogeneous engines \protect \greencirclednum{2} heterogeneous engines and \protect \bluecirclednum{3} co-designed engines (CMEA).}
     \label{fig:multi_engine_construction}
 \end{figure}

Multi-engine DL accelerators comprise independently controllable computing engines interconnected through a high-bandwidth communication fabric, where each engine mainly consists of an array of Processing Elements (PEs), as well as local and global buffers~\cite{gao2019tangram, kwon2021heterogeneous, zheng2022atomic, cai2023inter, shao2021simba, cai2024gemini, das2024multi}. These accelerators can be \textit{model-specific} or \textit{flexible}. Model-specific ones typically leverage reconfigurable hardware and are optimized for a single or a few very similar workloads (models)\cite{qararyah2025mcexplorer, hao2019fpga, shen2023mars}. In contrast, flexible accelerators are optimized for diverse DL workloads~\cite{gao2019tangram, kwon2021heterogeneous, cai2024gemini}.
We classify existing flexible multi-engine DL accelerators, depending on engine heterogeneity, into the two categories depicted in Figure~\ref{fig:multi_engine_construction}: \protect \redcirclednum{1} homogeneous accelerators, which use the same hardware configurations across all engines~\cite{shao2021simba, cai2024gemini}; and \protect \greencirclednum{2} heterogeneous accelerators, in which engines differ at the architectural or microarchitectural level~\cite{kwon2021heterogeneous, das2024multi, odema2024scar}.

Existing flexible multi-engine DL accelerator design approaches combine \emph{independently optimized expert-selected} engines such as Eyeriss~\cite{chen2016eyeriss, chen2019eyeriss}, ShiDiaNao~\cite{du2015shidiannao}, or NVDLA~\cite{nvdla}. These approaches exhibit several limitations. \textbf{First}, although expert-driven selection can produce optimized accelerators for particular scenarios, \emph{it lacks a robust and reusable methodology}. A reusable methodology is needed as the fast pace of DL model evolution requires continuous accelerator co-design to sustain performance and efficiency~\cite{sakhuja2023leveraging}. Furthermore, DL workloads span a broad range of deployment environments and resource constraints; designs tailored for edge settings may not translate effectively to cloud-scale systems~\cite{kwon2021heterogeneous}. 
\textbf{Second}, due to the exponentially large DL accelerator design space, \emph{systematic methodologies consistently outperform expert-driven ones}, as demonstrated in single-engine design where improvements range from two times to an order of magnitude~\cite{dave2023explainable, zhang2022full, parashar2019timeloop, kao2022digamma, sakhuja2023leveraging, hong2023dosa}. \textbf{Third}, independently designed engines may become \emph{optimized for the same or substantially overlapping subsets of workloads}, resulting in poor coverage of the overall workload landscape. We define workload \emph{coverage} as an accelerator's ability to process a set of workloads with minimal aggregate execution cost (such as execution time or energy consumption).
 
To overcome these limitations, this work proposes \underline{M}ulti-\underline{E}ngine DL accelerator \underline{DE}sign \underline{M}ethodology (MEDEM). The idea behind MEDEM is to decouple engine co-design and selection as highlighted in Figure~\ref{fig:multi_engine_construction}. MEDEM's \textbf{first stage} (\protect \bluecirclednum{a}) aims to \textit{co-design a pool of candidate engines} that conceptually achieve a maximal coverage of diverse workloads, by co-designing an engine for each kernel with distinct computational characteristics. However, in practice, resource constraints often limit the number of engines that can be included. Hence, MEDEM's \textbf{second stage} (\protect \bluecirclednum{b}) \emph{selects a multiset} \footnote{A multiset is a collection of elements in which repetitions are allowed.} of engines (from the pool) that maximizes workload coverage given a practical resource budget. We call accelerators designed using this methodology \textbf{Co-designed Multi-Engine Accelerators (CMEAs)}. On the one hand, co-designing a relatively large pool of engines boosts the chance of identifying a combination of engines with complementary capabilities, hence achieving maximum coverage. On the other hand, it requires very efficient co-design and selection strategies to navigate exponentially large design spaces.

Engine co-design entails exploring two exponentially large and non-convex subspaces. These are the hardware space and the mapping space~\cite{hong2023dosa, hegde2021mind}. The state-of-the-art hardware-mapping co-design approaches use techniques ranging from simple to sophisticated DL-based ones~\cite{kao2022digamma, hong2023dosa, huang2022learning, hegde2021mind, xiao2021hasco, sakhuja2023leveraging, dave2023explainable}. However, as important as the techniques themselves, if not more, is effectively navigating the design space~\cite{xiao2021hasco, huang2022learning, dave2023explainable}. The engine co-design stage of MEDEM proposes a \textbf{Hierarchical Co-designer (HCo)} to navigate the hardware-mapping space effectively. The idea is to \emph{prioritize} certain axes of the space - such as PE arrangements or tiling - based on their \emph{impact} on the execution cost, regardless of whether they belong to the hardware or mapping space. As prioritized axes have a large impact, they are explored exhaustively, whereas lower-priority axes are explored heuristically. To enable exhaustive exploration, HCo proposes pruning techniques that use domain knowledge to eliminate redundant and provably suboptimal regions of the space.

\reviewtxtv{Identifying a combination of engines that achieve maximum workload coverage through exhaustive exploration is intractable. For example, selecting a multiset of 64 engines from 100 candidates requires exploring $\mathcal{O}(10^{46})$ possible combinations. To tackle this problem, we propose a strategy that balances two intuitive design principles: (1) designing for the common case by replicating the engine with the best average performance, and (2) maximizing specialization through engine heterogeneity. Rather than committing to either extreme, the strategy explores multiple engine combinations, drawn from the pool generated by MEDEM's first stage, that span the spectrum from one extreme to the other. Each combination is then incrementally refined, and the combination with the lowest aggregate execution cost is finally selected. We call this strategy \textbf{Balanced Average and Specialization Search (BASS)}.}

Our contributions are as follows:

\begin{itemize}
    \item We propose MEDEM, a systematic design methodology for flexible multi-engine DL accelerators. MEDEM is a two-stage methodology that decouples engine co-design and selection.
    \item We propose HCo, a co-design strategy that efficiently navigates an engine hardware–mapping space by exhaustively exploring high-impact axes and heuristically exploring the rest.
    \item \reviewtxtv{We propose BASS, a strategy for selecting engine combinations that maximizes workload-set coverage by exploring combinations ranging from ones designed for the common case to highly specialized ones.}
    \item We comprehensively evaluate MEDEM and its derived CMEAs against state-of-the-art homogeneous and heterogeneous multi-engine accelerator designs, demonstrating MEDEM's scalability and generalizability.
\end{itemize}

Comprehensive evaluation of MEDEM using 51 single- and multi-model workloads shows that it identifies CMEAs that outperform state-of-the-art homogeneous and heterogeneous multi-engine accelerator baselines modeled after Simba~\cite{shao2021simba} and Maelstrom~\cite{kwon2021heterogeneous}, achieving geometric mean improvements up to $4.84\times$ in energy-delay product (EDP) and $1.59\times$ in throughput. These improvements hold across various resource budgets ranging from those typical of edge to cloud systems.

\section{Background and Motivation}
\label{sec:back}
\subsection{Deep Learning Models and their Unique Layers}
\label{subsec:back_models}
\begin{figure}[htbp]
         \centering
         \includegraphics[width=0.9\columnwidth]{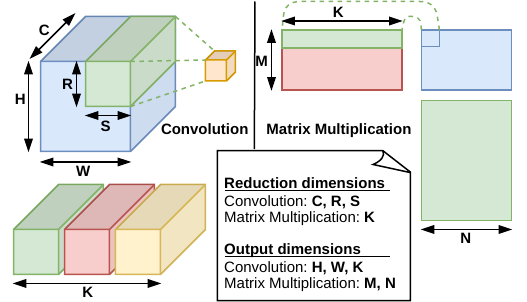}
         \caption{Convolution and matrix multiplication dimensions, reduction and output dimensions.}
         \label{fig:conv_matmul}
\end{figure}

Deep Learning (DL) models, such as Convolutional Neural Networks (CNNs)~\cite{lecun2015deep} and Transformers~\cite{vaswani2017attention}, consist of a sequence of layers. Each layer performs a specific tensor operation to transform input data into alternative feature representations. Two core operations of these DL models, which are characterized by their multi-dimensional iteration spaces, are convolution and matrix multiplication~\cite{kwon2020flexion, parashar2019timeloop}. These operations fundamentally combine \emph{input} activations with learned \emph{weights} to produce output activations. A standard convolutional layer operates within a 6-dimensional space (K, C, H, W, R, S) shown in Figure~\ref{fig:conv_matmul}, while a matrix multiplication operation has a 3-dimensional space (M, N, K), assuming no batching. We define a \emph{unique layer} as a layer that has a specific set of values for all its dimensions. \emph{Modern DL models, despite having tens to thousands of layers, are often composed of only a few unique layers that repeat throughout the model.} We refer to layers sharing identical dimensional values as \emph{instances or repetitions} of a single unique layer.

\subsection{DL Computing Engines}
\label{subsec:back_engines}
We define an engine as an array of Processing Elements (PEs) with a multilevel memory hierarchy, including private register files (RFs), local and global buffers. Engines can be broadly categorized based on which dimensions of the DL layers can be mapped in parallel (i.e., spatially) across the PE array. Different dimensions of the core deep learning operations have two main forms of parallelism. First, parallelism on the output dimensions: (K, H, W) in convolution (Figure~\ref{fig:conv_matmul}) and (M, N) in matrix multiplication. We call parallelism on these dimensions \emph{non-reductive (NR)}. Second, parallelism on dimensions that require partial-sum accumulation: (C, R, S) in convolution and (K) in matrix multiplication, \emph{reductive (R)}. Supporting spatial mapping of these dimension types has implications for an accelerator microarchitecture and for data reuse opportunities~\cite{parashar2019timeloop, yang2020interstellar, jang2021sparsity, kwon2019understanding}. For example, when parallelizing on NR dimensions, each PE computes an output element independently and accumulates the result at its local register~\cite{du2015shidiannao}. Consequently, the \emph{hardware primitives} in the NR parallelism case do not require spatial reduction logic. By contrast, parallelizing R dimensions requires spatial reduction primitives, typically implemented via inter-PE adder trees or global accumulation registers~\cite{jang2021sparsity,shao2021simba}.

\begin{table}[htbp]
\small
\centering
\caption{Engine types used in existing multi-engine accelerators. NVDLA/Simba (\textbf{NV/Si}), Eyeriss (\textbf{Eye}), Systolic Array (\textbf{SA}), and ShiDianNao (\textbf{Shi}). \textbf{Het.} indicates heterogeneous engines. If an accelerator is homogeneous, multiple \cmark mean that different engines are used mutually exclusively, one type at a time, in different experiments.}
\label{tab:multi_eng_summary}
\begin{tabular}{|p{3.6cm}|c|c|c|c|c|}
\hline
\rowcolor{gray!10}
\textbf{Work} & \textbf{NV/Si} & \textbf{Eye} & \textbf{SA} & \textbf{Shi} & \textbf{Het.} \\
\hline
\raggedright Tangram~\cite{gao2019tangram}
&  & \cmark&  &  & \xmark \\
\hline
\raggedright AI-MT~\cite{baek2020multi}, Planaria~\cite{ghodrati2020planaria}, SPA~\cite{cai2022deepburning}, TPU v4~\cite{jouppi2023tpu}
&  &  & \cmark&  & \xmark \\
\hline
\raggedright Simba~\cite{shao2021simba}, NN-baton~\cite{tan2021nn}, SET~\cite{cai2023inter}, Gemini~\cite{cai2024gemini}
& \cmark&  &  &  & \xmark \\
\hline
\raggedright Atomic~\cite{zheng2022atomic}
& \cmark&  &  & \cmark & \xmark\\
\hline
\raggedright Crane~\cite{gong2025crane}
& \cmark&  & \cmark&  & \xmark\\
\hline
\raggedright ConfuciuX~\cite{kao2020confuciux}
& \cmark& \cmark&  & \cmark& \xmark\\
\hline
\raggedright MAGMA~\cite{kao2022magma}
& \cmark& \cmark&  &  & \cmark \\
\hline
\raggedright SCAR~\cite{odema2024scar}
& \cmark&  &  & \cmark& \cmark \\
\hline
\raggedright HDA~\cite{kwon2021heterogeneous}, MOHaM~\cite{das2024multi}
& \cmark& \cmark&  & \cmark& \cmark\\
\hline
\end{tabular}
\end{table}

\subsection{Multi-Engine DL Accelerators}
\label{subsec:back_multi}
Multi-engine DL accelerators consist of independently controllable engines interconnected via a high-bandwidth communication fabric. These accelerators offer several advantages over monolithic ones. \emph{First}, they offer more specialization opportunities by dedicating the engines to address the heterogeneity of DL model kernels, improving performance and efficiency~\cite{boroumand2021google, kwon2021heterogeneous, cai2022deepburning, zhang2018dnnbuilder, das2024multi}. \emph{Second}, they exploit model-level parallelism by enabling concurrent execution of different model layers~\cite{gao2019tangram, zheng2022atomic, cai2022deepburning, kwon2021heterogeneous}. \emph{Third}, their modular design improves scalability and flexibility for large models~\cite{shao2021simba, gong2025crane}. \emph{Finally}, they expose more mapping and scheduling optimizations for multi-model workloads~\cite{gong2025crane, odema2024scar, cai2024gemini, venieris2023multiple, verhelst2022ml}.

We define \emph{flexible multi-engine accelerator} as an accelerator capable of executing a diverse set of DL workloads, to distinguish from \emph{model-specific accelerators} usually optimized for specific DL models~\cite{qararyah2025mcexplorer, hao2019fpga}. Designing the engines of flexible multi-engine accelerators is largely expert-driven. As Table~\ref{tab:multi_eng_summary} shows, existing flexible multi-engine accelerators use combinations of independently optimized engine architectures.

\reviewtxt{\subsection{Motivating a Two-Stage Design Methodology}}

\reviewtxt{To fully exploit the potential of multi-engine DL accelerators, systematic design methodologies are needed. A natural extension of single-engine hardware-mapping co-design is to treat engine configurations and engine combinations as dimensions of a \emph{single joint design space} and explore them simultaneously. This approach is exemplified by MOSAIC~\cite{das2026mosaic}, the first framework to comprehensively target the design of flexible multi-engine DL accelerators. However, this approach has multiple shortcomings.}

\reviewtxt{\textbf{First}, the joint design space is dominated by combinations containing inefficient engines. This is because \emph{optimized configurations constitute only a small fraction} of all feasible designs in the per-engine design space~\cite{dave2023explainable,sakhuja2023leveraging,parashar2019timeloop,kao2020gamma}. Moreover, substantial quality variation exists even among top-performing engines; for example, TimeLoop reports up to an $11\times$ difference in energy efficiency among $6582$ designs that all achieve the minimum number of DRAM accesses~\cite{parashar2019timeloop}. Consequently, sampling the joint space is likely to spend most of its effort evaluating combinations composed of suboptimal engines. \textbf{Second}, jointly exploring engine configurations and engine combinations induces multiplicative growth in the search space, increasing exploration complexity and \emph{limiting scalability}. The practical impact of this growth can be observed in MOSAIC, whose evaluation limits the design space to accelerators with $ 24$ engines or fewer ($1$--$3$ engine types and $1$--$8$ instances per type~\cite{das2026mosaic}). \textbf{Third}, another impact of the space growth is practically \emph{limiting per-engine hardware-mapping co-design}. For instance, although MOSAIC explores $12$ design knobs, where DAG-aware mapping across heterogeneous engines is considered, within-engine mapping remains largely unexplored: loop ordering is not considered, and data tiling is restricted to values that fit the working set~\cite{das2026mosaic}. However, co-designing architectural and mapping dimensions is crucial for discovering highly efficient accelerator designs~\cite{kao2022digamma,sakhuja2023leveraging,hong2023dosa,xiao2021hasco,dave2023explainable,mei2021zigzag}.}

\reviewtxt{
These shortcomings motivate \emph{decoupling engine co-design and combination selection}. \textbf{First}, decoupling enables the identification of high-quality engine candidates, avoiding the expenditure of search effort on combinations involving poor or mediocre engines. \textbf{Second}, it enables a more structured and scalable exploration compared to simply exploring joint space. \textbf{Third}, as the decoupled (isolated) engine design space is smaller, co-designing engines by co-exploring the hardware and mapping spaces becomes more practical. Moreover, as the engine co-design problem is extensively studied in the literature, isolating it (the co-design) enables direct applications of insights from the rich literature.
}

\begin{figure*}[htbp]
     \centering
 \begin{subfigure}[t]{0.69\textwidth}
     \includegraphics[width=\textwidth]
     {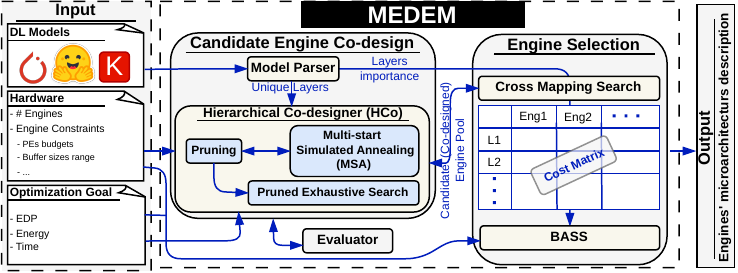}
     \caption{MEDEM overview. }
     \label{subfig:medem_overiew}
     \end{subfigure}
      \hfill
 \begin{subfigure}[t]{0.29\textwidth}
     \includegraphics[width=\textwidth]{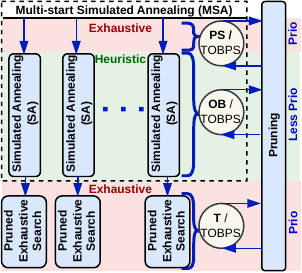}
     \caption{Hierarchical Co-designer (HCo) flow. }
     \label{subfig:medem_ho}
     \end{subfigure}
      \hfill
\caption{MEDEM overview and Hierarchical Co-designer flow.}
\label{fig:medem_overiew_and_ho}
\end{figure*}

\section{MEDEM Overview}
\label{sec:medem_overview}


\subsection{MEDEM Flow and Core Modules}
 \label{subsec:medem_flow}
 
Figure~\ref{subfig:medem_overiew} shows a high-level overview of the MEDEM flow and its core modules. MEDEM takes as \textbf{input} a set of DL models expressed using widely used DL frameworks, such as PyTorch, a hardware resource budget including the number of engines and the resource limit per engine, and an optimization goal. As an \textbf{output}, MEDEM produces hardware configurations of engines that form an optimized Co-designed Multi-Engine Accelerator (CMEA) and the number of instances of each engine. MEDEM has two main modules. The first is the \textbf{candidate engine co-design} module(Sections~\ref{sec:medem_candidate_gen}), which produces a pool of candidate engines, each optimized for a unique kernel (layer). The second is the \textbf{engine selection} module (Section~\ref{sec:medem_eingine_selection}), which selects a multiset of co-designed candidates that maximizes workload coverage, or minimizes aggregate execution costs with respect to the optimization goal. We quantify coverage using the \emph{geometric mean} of execution cost across the workload set. 

Although at runtime the workloads of a CMEA are complete DL models, at design time, MEDEM considers individual \emph{unique layers} (defined in Section~\ref{subsec:back_models}). MEDEM is \emph{model-topology-agnostic}, prioritizing flexibility for a diverse workload set. Considering model topologies introduces a trade-off between generality and the ability to maximize optimization opportunities for specific models. Since our goal is to design a \emph{flexible} accelerator for diverse models, MEDEM avoids optimizations tailored to specific model topologies. Instead, it aims to identify CMEAs that optimize aggregate execution cost across any set of models constructed from layers with given characteristics, independent of their topologies.

\begin{figure}[htbp]
         \centering
         \includegraphics[width=\columnwidth]{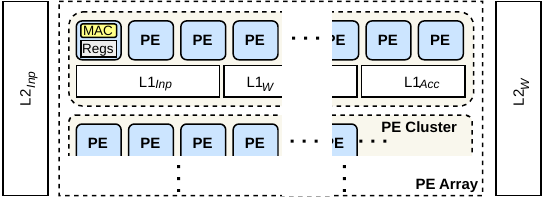}
         \caption{MEDEM engine abstraction, focusing on compute and memory components.}
         \label{fig:engine}
\end{figure}

\subsection{MEDEM Supported Engine Architectures}
\label{subsec:medem_eng_archs}
An engine is abstracted as a 2-D array of PEs exemplified in Figure~\ref{fig:engine}. Each PE contains a Multiply–Accumulate (MAC) unit and a small set of local registers. The PEs are grouped into 1-D PE clusters. A PE cluster has three buffers local to the cluster and accessible to all its PEs (L1 buffers). These buffers hold input, weights, and partial sums (accumulation results). The engine has two global buffers accessible by all its PE clusters, one for inputs and one for weights (L2 buffers). We model the buffers as buffets~\cite{pellauer2019buffets}. Buffets decouple fills and accesses through fine-grained synchronization, making them more efficient than the commonly used double-buffered scratchpads. We broadly categorize our engines into two types based on their hardware primitives support for parallelism on R and NR dimensions (Section~\ref{subsec:back_engines}). We call these two types \emph{R-engines} and \emph{NR-engines}. In the rest of the paper, we use the term \emph{unique engine} to refer to an engine that has a specific set of values for the type, PE array shape, and size of all buffers. \reviewtxt{Note that the R and NR are abstractions rather than two concrete engine configurations. This work explores instances of these abstractions with various PE array shapes, spatial mapping, and dataflow options, and buffer sizes. Moreover, given an engine instance, all valid loop orders and tiling options at each level of the memory hierarchy are explored.}
 
\section{Candidate Engine Co-design}
\label{sec:medem_candidate_gen}
This section presents MEDEM's first core module, the candidate engine co-design module, which comprises two components. First, \emph{model parser}, which extracts \emph{unique layers} (defined in Section~\ref{subsec:back_models}) from the input DL models. Second, \emph{Hierarchical Co-designer (HCo)}, which is the key component of this module. HCo co-designs a pool of engines given the unique layers and resource budgets. As the same co-designed engine can be optimal for multiple layers, the engine pool may have fewer engines than the unique layers.

\subsection*{Hierarchical Co-designer (HCo)}
\label{subsec:medem_ho}
The size and structure of an engine hardware-mapping co-design space make uniform exploration unlikely to produce consistently high-quality solutions. Hence, HCo uses hierarchical exploration \emph{prioritizing important axes} of the space based on their impact on the execution cost. The prioritized axes are explored exhaustively, and the rest are explored heuristically. Exhaustive exploration is enabled through domain-informed design space pruning.

\subsubsection{Importance-based Design Axes Prioritization} 

HCo's axes prioritization relies on a four-axis representation of the hardware-mapping space, named \emph{TOPS}~\cite{kao2022formalism}. TOPS stands for tiling (T), loop orders (O), parallelization (P), and PE array shapes (S). However, we separate the tiling and buffer sizes (B), naming this variant TOBPS. An extensive analysis of these axes suggests that optimizing \emph{parallelism and PE array shape (PS)} yields considerable performance gains~\cite{kao2022formalism}. The same applies for \emph{blocking or tiling (T)},~\cite{kao2022formalism, sakhuja2023leveraging, yang2020interstellar}. Hence, HCo prioritizes these axes, namely (PS) and (T), and explores them exhaustively, and explores (OB) heuristically.

\subsubsection{HCo Components and Flow}
Figure~\ref{subfig:medem_ho} shows HCo's flow and sub-optimizers. The first sub-optimizer is a Multi-Start Simulated Annealing (MSA). The MSA explores \emph{parallelism and PE array shape together (PS)} exhaustively. We apply simple pruning on array shape (S), limiting its dimensions to powers of two; with this, the P and S options are usually tens to hundreds, making an exhaustive exploration practical. Each SA search of the MSA explores the combinations of \emph{loop orders and buffer sizes (OB)} (heuristically). And for each explored combination, HCo starts an exhaustive search for \emph{optimal tile sizes and shapes (T)}. Unlike the (PS), the remaining axes have huge spaces and require considerable pruning. HCo's partitioning of the space and grouping of hardware and software axes together may sound counterintuitive. However, together with the applied pruning, it facilitates selective exploration of promising parts of the design space, leading to the identification of highly optimized design points (as demonstrated in Section~\ref{subsec:comp_with_spot}).

\subsubsection{Pruning Redundant, Dominated, and Undesired Design Points}
\label{subsec:medem_pruning}
Previous work primarily prunes \emph{invalid} design points, but the design space also contains valid yet suboptimal points. We classify these as \emph{redundant} (equivalent), \emph{dominated} (provably suboptimal), or \emph{undesired} (excluded due to designer preferences or implementation constraints). Because design-space axes are interdependent, pruning must be applied across all axes to enable exhaustive exploration of important ones. This section skips pruning invalid designs, which is well covered in the literature, and discusses the other three.

\textbf{Examples of redundant design points.} 
We define \emph{interchangeable dimensions} as dimensions (Section~\ref{subsec:back_models}) with identical values and input-to-output mappings. In the absence of parallelism on such dimensions, tiles (\textbf{T}) that differ only in the permutation of these dimensions' values provide equivalent PE utilization and reuse and are therefore redundant. Similarly, loop orders (\textbf{O}) that differ only in the permutations of not parallelized \emph{interchangeable dimensions} are redundant. Pruning redundant points shrinks the space considerably. For example, as our engines permit any loop order at any memory level (Section~\ref{subsec:medem_eng_archs}), there are $(7!)^3 \approx 12.8$ billion possibilities when considering convolution with batching. Interchangeable dimensions-based pruning can reduce these to just thousands.

\textbf{Examples of dominated design points.}
Orojenesis~\cite{huang2024mind} shows that, for matrix multiplication, the maximum effectual buffer size (\textbf{B}) is determined by the smallest operand (smallest operand size + its smallest rank + 1). We generalize this to convolution with an extra constraint: the smallest rank must be shared with the other input. Global buffers larger than the maximal effectual size are suboptimal (dominated) as they cost more area and energy per access with no added value. To give another example, we define two tiles (\textbf{T}) \textit{A} and \textit{B} as \textit{divisible}, with respect to a buffer size, if both fit in the buffer, \textit{A} is larger than \textit{B}, and each dimension of \textit{A} is an integer multiple of the corresponding dimension of \textit{B}. Any tile that divides \textit{A} is pruned as \emph{dominated}, since it is guaranteed to be suboptimal. This criterion is stricter than the common practice of retaining only the largest fitting tile. Because different dimensions may provide different reuse and PE utilization, a tile \textit{B} that is smaller than \textit{A}, but larger in at least one dimension, is not necessarily dominated by \textit{A}.

\textbf{Example of pruning undesired points.} We define \emph{collocated dimensions} as those that appear consecutively in a loop order (\textbf{O}). The idea of collocated dimensions is a \emph{generalization of the concept of stationary}~\cite{chen2016eyeriss} applied to different levels of the memory hierarchy. For example, any loop order where the dimensions of a layer output at a certain memory level are not collocated as outermost is not "output stationary" relative to that level. One can enforce stationary loop orders by pruning loop orders that have certain non-collocated dimensions.

\textbf{Gradual pruning relaxation.} Applying all pruning types may leave the design space with no valid solutions under the specified hardware budget. In this case, pruning is gradually relaxed by reintroducing previously pruned dominated and undesired designs, while redundant designs remain excluded because equivalent counterparts already exist in the pruned space. After each reintroduction, the exploration is repeated. Designs are added back in order of increasing design space expansion, meaning that pruning types with the smallest impact on the design space are relaxed first.

\begin{algorithm*}[htbp]
\small
\caption{Balanced Average and Specialization Search (BASS)}
\begin{algorithmic}[1]
\State \textbf{Input:} Engine Candidate Pool $\mathcal{P}$, Costs Matrix $C_{L \times E}$, Engine multiset Size $K$, Seeds Heterogeneity $H = \{1, 2, 4, \dots, K \}$ 
\State $engCombs \gets \emptyset$ \IDEComment{Storage for local optima combinations}
\For{$h \in H$} \IDEComment{Multiple searches: exploring the specialization spectrum}
    \State $C_h \gets \text{UniformReplicaInit}(\mathcal{P}, C_{L \times E}, K, h)$ \IDEComment{Seed a combination by replicating top-$h$ engines uniformly}
    \Repeat \IDEComment{Greedy Hill-Climbing refinement}
        \State $swapped, \delta \gets \text{bestOfAllSwaps}(C_h, \mathcal{P}, C_{L \times E})$ \IDEComment{Try swapping every engine in the combination}
        \State \textbf{if} $\delta > \epsilon$ \textbf{then} $C_h \gets swapped$ \IDEComment{Commit swap that improves by more than threshold $\epsilon$}
    \Until{$\delta \le \epsilon$} \IDEComment{Terminate at local optimum}
    \State $engCombs \gets engCombs \cup \{C_h\}$ \IDEComment{Add result to local optima set}
\EndFor
\State \Return $C \in engCombs$ with minimum weighted geometric mean \IDEComment{Return best overall combination (multiset)}
\end{algorithmic}
\label{algo:bass}
\end{algorithm*}

\section{Engine Selection}
\label{sec:medem_eingine_selection}
This section describes MEDEM's engine selection module, which identifies a multiset of co-designed engines that minimize an aggregate (measured as geometric mean) execution cost of workloads constructed from layers with given characteristics. Permitting multiple selections of the same engine enables the reuse of high-utility engines, which increases the probability that relatively more \emph{important layers} (Section~\ref{subsec:medem_metric_matrix}) are executed on engines that best suit their computational characteristics. The size of the search space of selecting combinations of $K$ engines out of $E$ candidates is shown in Equation~\ref{eq:space_size}. The engine selection module (Figure~\ref{subfig:medem_overiew}) contains two components. The \emph{cross mapping search} constructs a \emph{cost matrix} given the unique layers extracted from the input models and the co-designed engine pool, as described in Section~\ref{subsec:medem_metric_matrix}. The second is \emph{Balanced Average and Specialization Search (BASS)}, which selects the best combination (multiset) of candidate engines to construct a CMEA as described in Section~\ref{subsec:medem_msls}.

\begin{equation}
\footnotesize
\label{eq:space_size}
\binom{E + K - 1}{K} = \frac{(E + K - 1)!}{K!(E - 1)!}
\end{equation}

\subsection{cost matrix construction}
\label{subsec:medem_metric_matrix}
Selecting the best $K$ engines from $E$ candidates requires first evaluating the cost of executing every unique layer on every engine. For engines other than the one co-designed for a given layer (Section~\ref{sec:medem_candidate_gen}), this cost is not uniquely defined, as it depends on the mapping. This is the task of the \emph{cross mapping search} (Figure~\ref{subfig:medem_overiew}), which basically reuses the HCo to search for the mapping parameters, while fixing the hardware parameters. The execution costs of the best mappings are weighted by the importance of the unique layers and recorded in the \emph{cost matrix}. \reviewtxtv{We define \emph{layer importance} as the number of times instances of a layer appear across all input models. This definition does not account for the computational complexity of the layers, as the impact of complexity is already reflected in the obtained execution costs.\footnote{MEDEM also allows specifying custom importance values.}}

\subsection{Balanced Average and Specialization Search (BASS)}
\label{subsec:medem_msls}

\reviewtxtv{When designing multi-engine accelerators, two principles can be pursued: optimizing for the common case by replicating an engine with the best average execution cost and maximizing specialization through engine heterogeneity. BASS views these principles as opposing endpoints of a design spectrum and explores engine combinations spanning different points along that spectrum. The combinations are drawn from the candidate pool generated in MEDEM's first stage. The combinations are refined independently, and ultimately the combination with the lowest aggregate execution cost is selected.}

\reviewtxtv{
BASS is outlined in Algorithm~\ref{algo:bass}. The algorithm describes searching for a combination (multiset) of $K$ engines from a pool of $E$ candidates using a multi-start local search. Each local search starts \textit{at line 4} with an initial engine combination constructed by uniformly replicating the top $h$ engines ranked by average execution costs across all layers (obtained from the cost matrix in Section~\ref{subsec:medem_metric_matrix}). For example, when $h$ is $4$, and $K$ is $64$, each of the top $4$ engines is replicated $16$ times. Different initializations represent points on a spectrum from designing for the common case by replicating one engine ($h = 1$) that is best on average, to maximizing specialization or engine heterogeneity ($h = K$). Each initial combination is incrementally refined using a greedy hill-climbing search (lines 5–8). At each iteration, the algorithm evaluates all possible engine swaps—\textit{i.e.}, replacing each of the $K$ engines in the current combination with every candidate engine in $E$—and selects the swap that provides the greatest marginal gain (line 6). The gain is defined as the reduction in aggregate execution cost. The gain is a proxy indicating that the swap results in a combination that comprises engines with better complementary capabilities. If the gain exceeds a predefined threshold\footnote{In all experiments, the threshold was set to $\epsilon = 10^{-9}$ solely to avoid floating-point round-off errors and did not affect the optimization.}, the swap is accepted, and the combination is updated (line 7). Marginal gain is computed relative to the best combination found throughout the search. The process repeats until no swap yields an improvement above the threshold, at which point a local optimum is reached (line 8). This local optimum is then added to the set of candidate solutions (line 9). After exploring multiple points on the heterogeneity spectrum, the algorithm compares the best local optimum obtained from each search and returns the one with the best workload coverage, measured by the minimum weighted aggregate cost (line 11).}

\begin{figure}[htbp]
         \centering
         \includegraphics[width=\columnwidth]{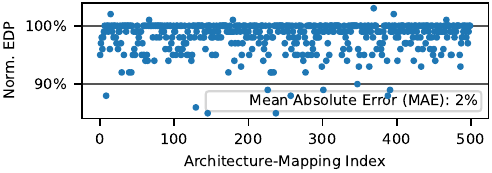}
         \caption{EDP estimated by our model normalized to Timeloop-Accelergy-Infrastructure using a random sample of 500 architecture-mapping.}
         \label{fig:accu_val}
\end{figure}

\subsection{Evaluator}
\label{subsec:medem_modeling}
Although pruning substantially reduces an engine hardware-mapping space, HCo may still explore billions of candidate configurations per layer, especially under relaxation. Hence, we develop a fast evaluation method. The core idea of our evaluator is similar to that of DOSA~\cite {hong2023dosa}. Namely, the evaluator is based on decoding Timeloop~\cite{parashar2019timeloop}, and converting part of the iterative program into mathematical models. As the evaluator is not a major contribution of this work, we highlight the main differences between our evaluator model and DOSA's and refer the reader to DOSA for detailed information\footnote{Moreover, MEDEM code, including the evaluator, will be open-sourced.}. \emph{First}, while DOSA supports only Systolic Arrays with square shapes and weight-stationary dataflow, our evaluator supports any engine that satisfies the abstraction described in Section~\ref{subsec:medem_eng_archs}. \emph{Second}, unlike DOSA, our modeling is not differentiable, as supporting a wider range of architectures makes it harder to formulate a differentiable model. The evaluator mainly estimates engine execution cycles and memory accesses at all levels of the memory hierarchy. To estimate energy, we collect energy per operation and energy per memory access from Timeloop-Accelergy infrastructure~\cite{parashar2019timeloop, wu2019accelergy}. The values are collected offline for all supported configurations of each hardware component. During co-design, the evaluator performs a lookup in the collected values. Figure~\ref{fig:accu_val} shows the EDP estimated by the evaluator compared to Timeloop's. As depicted, the average error margin is only $2\%$, and only a few cases have accuracy less than $90\%$.

\begin{table}[h]
\small
\centering
\caption{Used workloads. \textit{DT} stands for Design-Time.}
\begin{tabularx}{\columnwidth}{l|X} \hline
\multicolumn{2}{|l|}{\textbf{CNN Models}} \\ \hline
\textit{DT} & c1: VGG16~\cite{szegedy2016rethinking}, c2: VGG19, c3: InceptionV3~\cite{szegedy2016rethinking}, c4: Inception-ResNet-v2~\cite{szegedy2017inception}, c5: ResNet50~\cite{he2016deep}, c6: ResNet101, c7: MobileNet~\cite{howard2017mobilenets}, c8: MobileNetV2~\cite{sandler2018mobilenetv2}, c9: DenseNet121~\cite{huang2017densely}, c10: DenseNet201 \\ \hline
\textit{Unseen} & c11: ConvNeXt~\cite{liu2022convnet}, c12: ResNet-RS50~\cite{bello2021revisiting} \\ \hline

\multicolumn{2}{|l|}{\textbf{Transformer Models}} \\ \hline
\textit{DT} & t1: BigBird~\cite{zaheer2020big}, t2: DeBERTa~\cite{he2020deberta}, t3: BERT~\cite{devlin2019bert}, t4: RoBERTa~\cite{liu2019roberta}, t5: DistilBERT~\cite{sanh2019distilbert}, t6: ELECTRA~\cite{clark2020electra}, t7: ALBERT~\cite{lan2019albert}, t8: XLM-R~\cite{conneau2020unsupervised}, t9: T5, t10: GPT-Neo~\cite{black2022gpt}, t11: CamemBERT~\cite{martin2020camembert}, t12: XLNet~\cite{yang2019xlnet} \\ \hline
\textit{Unseen} & t13: ViT~\cite{dosovitskiy2020image}, t14: BART~\cite{lewis2020bart}, t15: Llama-3 (8B)~\cite{grattafiori2024llama} \\ \hline

\multicolumn{2}{|l|}{\textbf{Multi-Model}} \\ \hline
\textit{DT} & \textbf{2-model}: (c2,t8), (c4,t3), (c10,t11), (c1,t2), (c6,t7), (c3,t12), (c5,t9), (c8,t6), (c7,t5), (c9,t4) \newline \textbf{3-model}: (c2,c9,t12), (c6,c8,t3), (c1,c5,t2), (c10,c3,t9), (c7,c4,t4) \\ \hline
\textit{Unseen} & \textbf{2-model}: (t13,c12), (t13,c11), (t14,c11), (t14,c12), (t15,c12), (t15,c11) \newline \textbf{3-model}: (t13,c12,c11), (t14,c11,c12), (t15,c11,c12) \\ \hline
\end{tabularx}
\label{tab:models_summary}
\end{table}
\begin{table*}[htbp]
\small
\centering
\setlength{\tabcolsep}{1pt}    
\caption{The explored possibilities of key hardware and mapping parameters in the evaluation (per engine).}
\begin{tabularx}{\textwidth}{|p{1.4cm}|p{1.3cm}|p{1.3cm}|p{1.1cm}|p{1.2cm}|X|p{1.7cm}|X|} \hline
\rowcolor{gray!10}
\textbf{PE Array} & \textbf{Regs/PE} & \textbf{L1 $_{(I/W)}$} & \textbf{L1$_{Acc}$} & \textbf{L2$_{(I/W)}$} &\textbf{Parallelism and dataflow} & \textbf{Loop Order} & \textbf{Tiling} \\ \hline
X: 8--128 \newline Y: 8--128 & 
8--128 \newline (B) & 
2--32 \newline (KiB) & 
0.25--4 \newline (KiB) & 
64--1024 \newline (KiB) & 
All possible Convolution $\cap$ Matrix multiplication dimension mappings &
$(3!)^3$, $(6!)^3$ & 
Divisors of layer shape $\cup$ Multipliers of PE array shape \\ \hline
\end{tabularx}
\label{tab:dse_configs}
\end{table*}

\section{Evaluation}

\subsection{Experimental Setup}
\label{subsec:exp_setup}

\subsubsection{Baselines}
\label{subsec:baselines}
To demonstrate the efficiency of MEDEM's engine co-design stage, we compare HCo to Spotlight~\cite{sakhuja2023leveraging}, a state-of-the-art sample-efficient co-design tool. Spotlight originally uses MAESTRO~\cite{kwon2020maestro}. However, the authors also present TimeLoop-based evaluation and show considerable differences between the two in some cases. Hence, for fairness, we re-evaluate the top-10 design points suggested by Spotlight using our Timeloop-Accelergy-based evaluator (Section~\ref{subsec:medem_modeling}) and use the best among the results as our baseline.

We evaluate MEDEM-derived CMEAs against Simba-like~\cite{shao2021simba} (Simba-L), and Maelstrom-like~\cite{kwon2021heterogeneous} (Maelstrom-L) accelerators. Simba-L represents a homogeneous multi-engine accelerator, and Maelstrom-L represents a heterogeneous one. NVDLA-based multi-engine accelerators like Simba are the most common in related work~\cite{shao2021simba, cai2023inter, cai2024gemini,odema2024scar,gong2025crane, zheng2022atomic, tan2021nn}. Maelstrom heterogeneous engines use NVDLA and ShiDiaNao. To evaluate scalability, we evaluate engine counts from 4 ($2\times2$) to 64 ($8\times8$), representing a broad spectrum of resource budgets and deployment scenarios, from edge to cloud systems. \reviewtxt{As MEDEM is a design methodology, the goal is to compare its outcomes with accelerators that capture the key characteristics of Simba and Maelstrom, rather than specific implementations. However, for fairness, we consider two configurations for each baseline and compare with the one that gives the baseline the best results throughout the evaluation. In the first configuration, engines use hardware resources similar to those reported in the original papers. In the second, engine resources are scaled uniformly, up or down, so that the overall area is similar to the corresponding CMEA's.} 




\subsubsection{System Modeling}
\label{subsec:system_modeling}
The energy and execution time of the engines are modeled and validated against Timeloop~\cite{parashar2019timeloop} and Accelergy~\cite{wu2019accelergy} (which use Aladdin~\cite{shao2014aladdin} and CACTI~\cite{muralimanohar2009cacti} plug-ins under the hood) as described in Section~\ref {subsec:medem_modeling}. We use TimeLoop-Accelergy's 7nm technology node values in all our experiments, except for the comparison with Spotlight (Section~\ref{subsec:comp_with_spot}), where we use TimeLoop-Accelergy's 28nm values to match Spotlight's technology node. Regarding the inter-engine communication fabric, we model a mesh NoC using a customized implementation of the open-source NoC modeling in SET~\cite{cai2023inter}. The customization adds support for heterogeneous engines, which SET does not support natively. SET NoC modeling is used by previous work on multi-engine accelerators~\cite{gong2025crane, cai2024gemini}. The NoC bandwidth is set to 32 GB/s per link, and the energy per hop to 0.7 pJ/bit.

\subsubsection{Workloads}
\label{subsec:workloads}
We evaluate widely used CNN and Transformer models listed in Table~\ref{tab:models_summary}. The design-time (DT) models in the table are those whose unique layers are used at the design phase. The Unseen models are used to evaluate MEDEM generalization. The unseen models comprise both evolved versions of the DT ones and models exhibiting structural or paradigm shifts. We form multi-model scenarios (2-model and 3-model) by combining a Transformer with one or two CNNs. As Transformers are built from a few repeating unique layers, the goal of combining a Transformer with two CNNs is to create multi-model workloads where heterogeneity, in terms of unique layer counts in a workload, is maximized. As there are numerous combinations of the DT models, the multi-model DT workloads are formed randomly. For unseen models, we consider all possible combinations. The evaluation uses the commonly used batch size of $64$. \reviewtxt {The weights and activations use int8 quantization, and the accumulators use int32. In this work, we assume that off-chip memory capacity is not a bottleneck, meaning that it can hold the weights of large models like Llama-3 (8B).}

\subsubsection{Workload Mapping Across Engines}
\label{model_scheduling}
We use two common heuristics to map a workload on the engines of a multi-engine accelerator, namely layer-sequential (LS) and layer-pipeline (LP)~\cite{kao2020confuciux, cai2023inter, zheng2022atomic}. LS processes layers one at a time, potentially using all resources per layer. LP distributes the layers across the engines and pipelines their execution. Both heuristics are applied, and the results of the best among them are adopted both for CMEAs and for the baselines.

\subsubsection{Design Space and MEDEM Parameters}
Table~\ref{tab:dse_configs} summarizes the key per-engine parameters explored by the co-design stage of MEDEM. We avoid extremely skewed PE\footnote{The term PE is used differently in the literature. For example, the Simba paper uses it to refer to what corresponds to a PE cluster in our design.} arrays with a dimension smaller than $8$ as they do not perform well on average.
Regarding parallelism and dataflow, we explore all options common between convolution and matrix multiplication. This is because our accelerator must be flexible, where any layer must be executable on any engine. We explore all loop orders at the three levels of the memory hierarchy. This means up to $6!$ possibilities (Section~\ref{subsec:back_models}) per memory level for convolution and $3!$ for matrix multiplication.

The Simulated Annealing (SA) optimizer, used in HCo, employs $50$ main (cooling) steps with $100$ iterations per step (temperature). The SA initial temperature $T_0$ is dynamically estimated based on a random walk over $100$ samples from the space. An exponential cooling schedule is used with $\alpha = 0.95$. \internalcomment{Stochastic-algorithm variance}\reviewtxt{As SA is non-deterministic, we repeated each experiment 10 times using different random seeds. The observed variance on the aggregate results is negligible.}

\subsubsection{Evaluation Metrics and Granularity} We mainly evaluate EDP, as is common in related work~\cite{cai2023inter, hegde2021mind, kwon2021heterogeneous, sakhuja2023leveraging, huang2022learning, hong2023dosa, odema2024scar, gong2025crane}. We also present representative throughput results. We use the Geometric Mean (GM) to report aggregate performance across various workloads.

\begin{figure*}[htbp]
 \centering
 \begin{subfigure}[t]{0.46\textwidth}
         \centering
         \includegraphics[width=\textwidth]{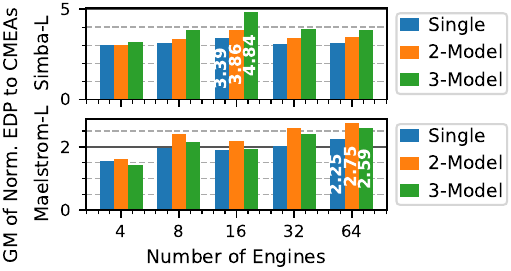}
         \caption{Geometric mean of baselines EDP normalized to CMEAs'.}
         \label{subfig:simba_comp_edp}
     \end{subfigure}
      \hfill
\begin{subfigure}[t]{0.46\textwidth}
         \centering
         \includegraphics[width=\textwidth]{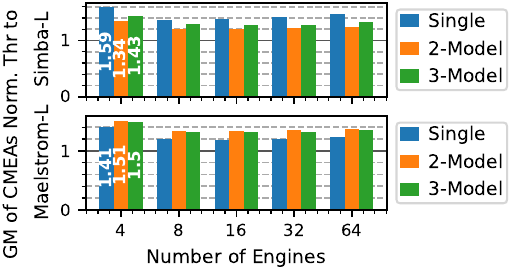}
         \caption{Geometric mean of CMEAs' throughput normalized to baselines.}
         \label{subfig:simba_comp_thr}
     \end{subfigure}
     \hfill 
\caption{EDP and throughput comparison of MDEDEM-derived CMEAs to baselines using DT models using single- and multi-model workloads. Numbers in bars indicate the best case for a workload set.}
\label{fig:simba_comp}
\end{figure*}

\subsection{End-to-End Evaluation of MEDEM}
\subsubsection{Evaluation of MEDEM-Derived CMEAs}
\label{subsec:comp_to_simba}

This section evaluates MEDEM-derived CMEAs compared to homogeneous multi-engine accelerators (Simba-L instances) and heterogeneous ones (Maelstrom-L instances).

Figure~\ref{fig:simba_comp} shows the EDP and throughput improvements achieved by CMEAs over Simba-L and Maelstrom-L. Compared to Simba-L, CMEAs achieve up to $4.84\times$ lower EDP and $1.59\times$ higher throughput. Compared to Maelstrom-L, CMEAs achieve up to $2.75\times$ lower EDP and $1.51\times$ higher throughput. Improvements in throughput are more modest than those in EDP. This is because optimizing throughput requires only maintaining full PE utilization, which is achievable for many layers by properly mapping a layer's dimensions on the PE array and by overlapping computation with data movement. In contrast, optimizing energy is more challenging as it requires maximizing spatial and temporal reuse across the memory hierarchy to reduce data movement costs. This makes EDP optimization more challenging; therefore, the remainder of the evaluation focuses on EDP.

Comparing results across different \emph{engine counts} shows that CMEAs consistently outperform baselines at all engine counts. On the one hand, consistent improvements highlight the \emph{MEDEM engine selection scalability}, as the design space of possible engine combinations grows exponentially with engine count. On the other hand, one would expect CMEAs' improvements over baselines to grow with engine count as CMEAs can be more heterogeneous. In other words, while Simba-L and Maelstrom-L replicate one or two engine architectures, respectively, MEDEM can specialize to better capture workload heterogeneity. This expected trend does not always materialize due to the interplay of \emph{two factors}: the limited heterogeneity of some models and the imbalance in layer contributions to the overall model computations. The impact of these factors is described in detail in Section~\ref{subsec:eval_bass}. Comparing improvements for~\emph{single- and multi-model} workloads in Figure~\ref{fig:simba_comp} shows mixed trends. While improvements over Simba-L increase with the number of models per workload, the largest improvements over Maelstrom-L are observed using 2-model workloads. Again, the two mentioned factors help to explain the trends in these results as detailed in Section~\ref{subsec:eval_bass}.

Overall, the fact that CMEAs improve over baselines across all experiments using representative workloads and at different scales demonstrates that \emph{constructing multi-engine accelerators using co-designed engines rather than independently optimized ones achieves better workload coverage, \textit{i.e.}, minimizes aggregate execution cost}. 

\begin{figure}[htbp]
\centering
     \includegraphics[width=\columnwidth]{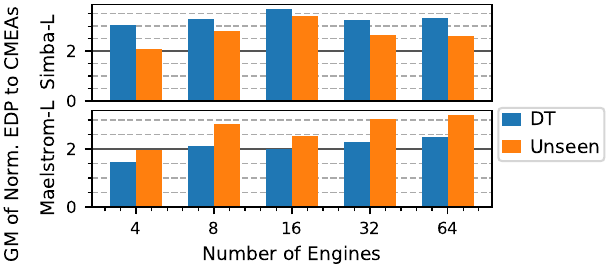}
\caption{Comparison between improvements using design-time (DT) and unseen models. Each bar is the geometric mean (GM) of all the DT/unseen workloads, \textit{i.e.} GM(single workloads, 2-model workloads, and 3-model workloads).}
\label{fig:unseen_comp}
\end{figure}

\subsubsection{Evaluation of MEDEM Generalizability}
\label{subsec:eval_unseen}
Figure~\ref{fig:unseen_comp} compares EDP improvements over baselines for both DT and unseen workloads. MEDEM-derived CMEAs outperform baselines using unseen workloads across all engine counts, suggesting that CMEAs remain effective beyond the workloads whose layers were used during design-time. While CMEAs' improvements over Simba-L decrease for unseen compared to DT workloads, improvements over Maelstrom-L increase. When considering both baselines together, the opposing trends largely balance out. The overall trends indicate that \emph{MEDEM generalizes relatively well}. As discussed in Section~\ref{subsec:medem_flow}, MEDEM is model-agnostic; it relies solely on the overall distribution of unique layers across a representative set of workloads, rather than a workload topology. Hence, as long as the design-time workloads are representative of the DL workloads' landscape, the aggregate results of MEDEM-derived CMEAs should be generalizable.

\begin{figure}[htbp]
         \centering
         \includegraphics[width=\columnwidth]{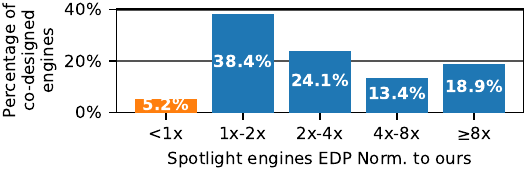}
         \caption{Histogram of EDP achieved by engines co-designed using Spotlight for all unique layers of the DT models, normalized to those co-designed using HCo. For example, $24.1\%$ of engines co-designed using HCo have $2\times$--$4\times$ EDP improvements over to those co-designed using Spotlight.}
         \label{fig:spotlight_comp}
\end{figure}

\subsection{Evaluation of MEDEM Engine Co-design Stage}

\subsubsection{Evaluation of HCo}
\label{subsec:comp_with_spot}
This section demonstrates the effectiveness of MEDEM's HCo by comparing its co-designed engines with those co-designed using the Spotlight framework~\cite{sakhuja2023leveraging}. In this evaluation, we first ran MEDEM's candidate engine co-design module to co-design engines for the unique layers of all DT models, as described in Section~\ref{sec:medem_candidate_gen}. Then we replace HCo with Spotlight, use this modified setup to co-design the engines, and then compare the EDP values achieved by engines co-designed using each approach to the other. Figure~\ref{fig:spotlight_comp} shows that HCo co-designed engines outperform Spotlight's in $94.8\%$ of the experiments, while Spotlight outperforms HCo only in $5.2\%$. Moreover, HCo identifies engines with considerable improvements that exceed $8\times$ compared to Spotlight in roughly one-fifth of the experiments. This is despite HCo being simpler than Spotlight, which relies on an expert-formulated sophisticated feature space~\cite{sakhuja2023leveraging}. These results demonstrate \emph{the effectiveness of importance-based prioritization of high-impact axes of the design space}.

\subsubsection{The Impact of Engine Co-design}
\begin{figure}[htbp]
         \centering
         \includegraphics[width=0.95\columnwidth]{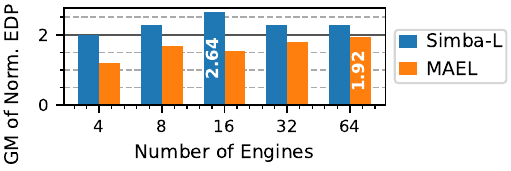}
         \caption{Comparsion of EDP achieved by CMEAs with homogeneous co-designed engines to the baselines (values are normalized to EDP of CMEAs with homogeneous co-designed engines). Each bar is the geometric mean (GM) of all 51 single- and multi-model DT and unseen workloads.}
         \label{fig:co_design_effect}
\end{figure}

This section focuses on isolating the impact of engine co-design on the efficiency of a multi-engine accelerator by comparing CMEAs with homogeneous engines to Simba-L and Maelstrom-L. These homogeneous CMEAs follow the principle of design for the common case by replicating the engine with the best average execution cost from the engine pool (Section~\ref{sec:medem_candidate_gen}). Figure~\ref{fig:co_design_effect} shows improvements over both Simba-L and Maelstrom-L of up to $2.64\times$ and $1.92\times$, respectively. These results emphasize the \emph{importance of co-designing the engines of a multi-engine accelerator}, even when they are homogeneous.

\begin{figure}[htbp]
         \centering
         \includegraphics[width=0.95\columnwidth]{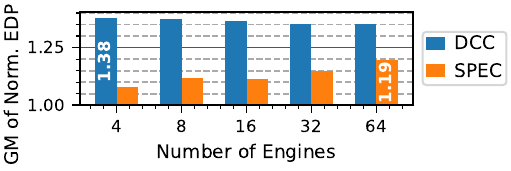}
         \caption{Comparison of BASS to two design alternatives: designing for common case (DCC), and maximizing specialization (SPEC). The values are normalized to BASS's. Each bar is the geometric mean (GM) of all 51 single- and multi-model DT and unseen workloads. DCC replicates one engine with the best average execution cost across all DT workloads K times (where K is the number of engines). SPEC comprises top-K unique engines ranked by their execution costs.}
         \label{fig:bass_eval}
\end{figure}

\subsection{Evaluation of MEDEM Engine Selection Stage}
\label{subsec:eval_bass}

This section isolates the impact of MEDEM's engine selection by comparing BASS with two alternative design principles: designing for the common case (DCC) and maximizing specialization (SPEC). Figure~\ref{fig:bass_eval} presents this comparison. BASS identifies engine combinations that outperform both alternatives across all engine counts, demonstrating its effectiveness. Furthermore, while improvements over DCC remain relatively stable, improvements over SPEC increase with the engine count. From another angle, the performance gap between SPEC and DCC narrows as the engine count increases, suggesting that the benefits of engine heterogeneity exhibit diminishing returns. Explaining these trends requires considering two factors together: the \emph{number of unique layers} in a workload, which is a proxy of that workload heterogeneity; and the \emph{contribution} of these unique layers' instances to the overall workload MAC operations.

\begin{figure}[htbp]
         \centering
         \includegraphics[width=\columnwidth]{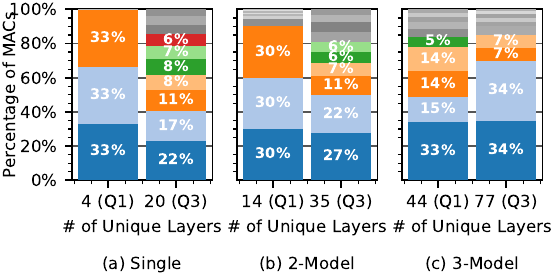}
         \caption{Unique layers and their contributions to the total MACs of a workload in single and multi-model workloads. A unique layer's contribution is calculated by aggregating MACs across all its instances (\textit{i.e.} repetitions). A sample of two workloads is shown per workload set, representing the first (Q1) and third (Q3) quartiles of workloads sorted by their unique layer counts. Unlabeled gray slices in a bar represent layers whose overall instances contribute less than $5\%$.}
         \label{fig:unique_layers_contribs}
\end{figure} 

Figure~\ref{fig:unique_layers_contribs} depicts representative statistics of the two factors mentioned. Using single-model workloads as an example, Figure~\ref{fig:unique_layers_contribs} (a) shows two representative models: one from the first quartile (Q1) and one from the third quartile (Q3) of the models sorted by the number of unique layers. The Q1 model represents relatively homogeneous models with few unique layers. It consists mainly of instances (repetitions) of only four unique core layers. Three of the four layers contribute $99\%$ to the model's total MACs. Consequently, using instances of more than four heterogeneous engines would exceed the workload's inherent heterogeneity and provide no additional benefit. In contrast, the Q3 model contains 20 unique layers. Nevertheless, only a small subset of unique layers (more precisely, all instances of these unique layers) accounts for the majority of the workload's MAC operations. Considering Q1 and Q3 models together suggests a sweet spot in engine heterogeneity, beyond which additional specialization yields diminishing returns or even degradation. Heterogeneous workloads such as Q3 benefit from specialized engines, but the gains diminish as new engine types target layers contributing fewer MACs. In contrast, Q1-like workloads may be harmed by further specialization. In other words, as the total number of engines is fixed, introducing additional engine types to improve support for the long tail of layers present in Q3-like workloads comes at the cost of reducing the replication of engines that are more optimized for the instances of the few dominant layers in Q1-like workloads. Figures~\ref{fig:unique_layers_contribs} (b) and (c) show that similar trade-offs exist in multi-model workloads; even though there are more heterogeneous layers in a workload, few of them still dominate the overall computations. 

\begin{figure}[htbp]
         \centering
         \includegraphics[width=\columnwidth]{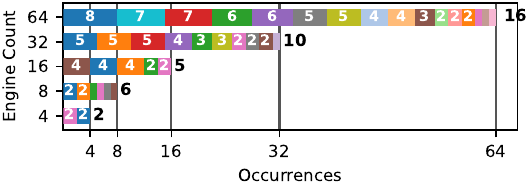}
         \caption{Engine combinations (multisets) selected by BASS and the contribution of unique engines. The number at the end of each bar indicates the number of unique engines in the selected combination. For example, for an engine count of $8$, BASS selects a multiset of $6$ unique engines, two of which are replicated twice, while the remaining four appear once. Colors encode engine configurations; slices with the same color across different bars correspond to the same engine configurations.}
         \label{fig:unique_engines_stack}
\end{figure}

Given such trade-offs, maintaining consistent improvements over DCC, across engine counts, indicates that \emph{MEDEM's engine selection avoids the trap of overspecialization}. Figure~\ref{fig:unique_engines_stack} shows the engine combinations selected by BASS across different engine counts. As observed, the number of unique engines is lower than the total engine count, suggesting diminishing returns from further specialization.

\begin{figure}[htbp]
         \centering
         \includegraphics[width=\columnwidth]{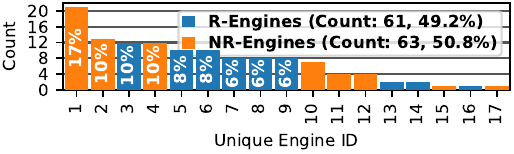}
         \caption{Occurrences of unique engines used across all CMEAs optimized for EDP. Percentages in bars are normalized to the total number of engines ($4 +8 + 16 + 32 + 64$).}
         \label{fig:engine_stats}
\end{figure}

\subsection{Overview of CMEAs Co-designed Engines}
\label{subsec:ceas_engines}
This section presents statistics and key architectural configurations of the engines selected by MEDEM. Figure~\ref{fig:engine_stats} shows that $17$ unique engines were selected across all experiments, with the most used engine selected $17\%$ of the time. Both R and NR engines have similar overall usage ($49.2\%:50.8\%$). Table~\ref{tab:top5_eng_archs} shows some hardware configurations of the most used $5$ engines. Different PE array aspect ratios are used. Various $L1$ and $L2$ buffer sizes are used with different input-to-weight buffer size ratios. This reflects the varying input-to-weight ratios among layers, which is a feature that must be captured by the buffer distribution to minimize off-chip accesses~\cite{qararyah2025mcexplorer}.

\begin{table}[h]
\footnotesize
\caption{Key hardware configurations of the Top 5 most used engines across all the CEAs optimized for EDP.}
\label{tab:top5_eng_archs}
\centering
\setlength{\tabcolsep}{2pt}
\begin{tabular}{|l|ccccccc|}
\hline
\rowcolor{gray!10}
Type & Percentage (\%) & PE Array & $L1_{Inp}$ & $L1_{W}$ & $L1_{Acc}$ & $L2_{Inp}$ & $L2_{W}$ \\
\hline
\rowcolor{plt_2_color!50}
NR & 17 & $32\times32$ & 2KiB & 2KiB & 1KiB & 128KiB & 128KiB \\
\rowcolor{plt_2_color!50}
NR & 10 & $128\times8$ & 8KiB & 2KiB & 1KiB & 1MiB & 512KiB \\
\rowcolor{plt_1_color!50}
R & 10 & $64\times16$ & 8KiB & 2KiB & 1KiB & 128KiB & 128KiB \\
\rowcolor{plt_2_color!50}
NR & 10 & $128\times8$ & 4KiB & 16KiB & 2KiB & 64KiB & 128KiB \\
\rowcolor{plt_1_color!50}
R & 8 & $64\times16$ & 2KiB & 2KiB & 256B & 64KiB & 128KiB \\
\hline
\end{tabular}
\end{table}

\section{Related work}

\textbf{DL engines hardware-mapping co-design.} Co-design of DL engines hardware–mapping is a widely explored open problem ~\cite{venkatesan2019magnet, yang2020interstellar, kao2020gamma, huang2021cosa, xiao2021hasco, kao2022digamma, sakhuja2023leveraging, hong2023dosa, dave2023explainable, kao2022formalism}. Existing work uses a various exploration heuristics including Genetic Algorithms~\cite{kao2020gamma, kao2022digamma}, Bayesian Optimization (BO)~\cite{xiao2021hasco, sakhuja2023leveraging}, Q-learning~\cite{xiao2021hasco}, variational autoencoders (VAEs)~\cite{huang2022learning}, and gradient descent~\cite{hong2023dosa}. The large size of the hardware and mapping spaces and the complex interactions between them necessitate sample-efficient exploration techniques~\cite{hong2023dosa, sakhuja2023leveraging, dave2023explainable}.

\textbf{Engines of flexible multi-engine DL accelerators.} Existing multi-engine DL accelerators comprise homogeneous engines~\cite{gao2019tangram, zheng2022atomic, zhang2022full, shao2021simba, cai2024gemini, jouppi2023tpu}, or heterogeneous ones~\cite{kwon2021heterogeneous, das2024multi, odema2024scar}. For example, Simba~\cite{shao2021simba} and Gemini~\cite{cai2024gemini} use NVDLA-like architecture in all their engines. SPA~\cite{cai2022deepburning} engines are dataﬂow-hybrid SAs that support Weight Stationary (WS) and Output Stationary (OS) dataﬂow. Atomic~\cite{zheng2022atomic} presents two multi-engine accelerator instances, one using NVDLA-like engines and one using ShiDiaNao-like~\cite{du2015shidiannao}. Maelstrom~\cite{kwon2021heterogeneous} engines use NVDLA and ShiDiaNao. MOHam~\cite{das2024multi} engines use NVDLA, Eyeriss, and ShiDiaNao. Another viable option is to use Reconfigurable Dataflow Accelerators (RDA), like MAERI~\cite{kwon2018maeri}. However, the literature shows that the overheads of RDA outweigh the gains~\cite {kwon2020flexion, kao2022formalism, kwon2021heterogeneous}.

\textbf{Model-specific multi-engine DL accelerators.} Model-specific, multi-engine DL accelerators usually use reconfigurable platforms. Reconfigurability facilitates dedicating the compute engines for a targeted model~\cite{zhang2018dnnbuilder,wei2018tgpa, blott2018finn, venieris2018fpgaconvnet, hao2019fpga}.  \reviewtxt{MCExplorer~\cite{qararyah2025mcexplorer} is a comprehensive DSE framework for model-specific multi-engine accelerators. The model-specific accelerator design problem has a smaller and simpler design space as each engine is tailored to one or a few layers, with many mapping parameters fixed in hardware.}

\reviewtxt{
\textbf{Comprehensive design methodologies of flexible multi-engine DL accelerators.} MOSAIC~\cite{das2026mosaic} is unique in providing a comprehensive exploration framework for flexible multi-engine DL accelerators, supporting a broad range of both MAC and non-MAC layers. However, MOSAIC jointly explores the engine-configuration and engine-composition spaces in an unstructured manner, limiting both effectiveness and scalability.}

Unlike existing flexible multi-engine DL accelerator design approaches, MEDEM proposes an efficient and scalable systematic methodology that decouples engine co-design from combination selection.

\section{Conclusion}
\label{sec:conclusion}
As monolithic DL accelerator optimizations reach diminishing returns, researchers are shifting toward multi-engine accelerators. This shift opens new design opportunities, but also introduces new challenges. A key challenge when designing a flexible multi-engine accelerator is identifying a combination of engines that maximizes coverage of diverse DL workloads to reduce aggregate execution costs. To tackle this challenge, this work proposes a systematic two-stage \underline{M}ulti-\underline{E}ngine DL accelerator \underline{DE}sign \underline{M}ethodology (MEDEM). Using generic engine abstractions, MEDEM co-designs candidate instances and selects a combination of the co-designed engines to improve coverage of diverse DL workloads. MEDEM encapsulates optimization strategies that efficiently navigate exponentially large design spaces, identifying flexible multi-engine accelerators that achieve geometric mean gains of up to $4.84\times$ in EDP and $1.59\times$ in throughput compared to the state-of-the-art. These gains are achieved using a representative set of DL workloads and under varying resource budgets.



\bibliographystyle{IEEEtranS}
\bstctlcite{IEEEexample:BSTcontrol}
\bibliography{references}
\end{document}